\documentclass{edbk} % Computer Modern font calls
\usepackage{epsfig}

\def\lap{\hbox{${_{\displaystyle<}\atop^{\displaystyle\sim}}$}}
\def\gap{\hbox{${_{\displaystyle>}\atop^{\displaystyle\sim}}$}}

\let\footnote\savefootnote

\let\footnotetext\savefootnotetext

\let\footnoterule\savefootnoterule

\normallatexbib
\begin{document}

%------------ article title  ------------------->>

% If you use \\'s , please supply an alternate version of the title
% in square brackets, i.e.,
%\articletitle[Communism, Sparta, and Plato]
%{COMMUNISM, SPARTA,\\ and PLATO}

\articletitle[Starquake-Induced Glitches]{STARQUAKE-INDUCED GLITCHES \\  IN
PULSARS}

%% optional, to supply a shorter version of the title for the running head:
%%\chaptitlerunninghead{}

\author{Richard I. Epstein}
\affil{Los Alamos National Laboratory}
\email{epstein@lanl.gov}

\author{Bennett Link}
\affil{Montana State University\\
Los Alamos National Laboratory}
\email{blink@dante.physics.montana.edu}

%% multiple authors may be separated with \\
%% \author{Samuel Bostaph\\
%% and Gregor Kariotis}

% optional prologue
%\prologue{<text>}{<author, year>}

% optional abstract
\begin{abstract}

\noindent The neutron star crust is rigid material floating on a
neutron-proton liquid core. As the star's spin rate slows, the
changing stellar shape stresses the crust and causes fractures. These
{\it starquakes} may trigger pulsar glitches as well as the jumps in
spin-down rate that are observed to persist after some
glitches. Earlier studies found that starquakes in spinning-down
neutron stars push matter toward the magnetic poles, causing
temporary misalignment of the star's spin and angular momentum.
After the star relaxes to a new equilibrium orientation, the magnetic
poles are closer to the equator, and the magnetic braking torque is
increased. The magnitude and sign of the predicted torque changes are
in agreement with the observed persistent spin-down offsets. Here we
examine the relaxation processes by which the new equilibrium
orientation is reached.  We find that the neutron superfluid in the
inner crust slows as the star's spin realigns with the angular
momentum, causing the crust to spin more rapidly.  For plausible
parameters the time scale and the magnitude of the crust's spin up
agree with the giant glitches in the Vela and other pulsars.

\end{abstract}

% optional keywords
% \begin{keywords}
% Text, text...
% \end{keywords}

%------------ body of article ------------------->>

\section{Introduction}

Stresses in the crust of a neutron star could produce starquakes that
 affect the star's spin evolution and generate high-energy
 emission. As the star's spin rate increases or decreases, changes in
 the equilibrium shape of the star and the differential rotation
 between the crust and the interior neutron superfluid generate
stress \cite{ruderman76}. In ``magnetars'', decay of the superstrong field
 ($B\gap 10^{14}$ G) could break the crust and drive episodes of
 intense gamma-ray emission \cite{tdtb}.  Recent studies showed that
 starquakes can change the magnetic spin-down torque acting on the
 star \cite{lfe98, fle}. Starquakes in slowing neutron stars drive
 matter toward the magnetic poles, distort the star's shape, and
 excite precession. As the precession damps, the star relaxes to a new
 rotational state with the magnetic  poles closer to the
 equator.  The new magnetic orientation enhances the braking torque
 on the star and may provide an explanation for the observed increases
 in the spin-down torque following glitches in the Crab pulsar,
 PSR1830-08 and PSR0355+54.

Here we investigate the physical processes that allow the star to
relax to its post-starquake equilibrium.  The most important processes
are the coupling between the liquid core and solid crust and the creep
of neutron superfluid vortices in the inner crust of the star. We find
that the changes produced by large starquakes can trigger catastrophic
unpinning of neutron superfluid vortex lines in the star's inner crust
\cite{andersonitoh}. As vortices move, the inner crust superfluid
rapidly settles to a state of lower angular momentum, while exerting a
spin-up torque on the crust.  Our estimates show that
starquake-triggered events may explain giant pulsar glitches as
well as the persistent spin-down offsets. The next section summarizes
earlier work on starquakes and spin-down offsets, and the following
one describes the post-starquake spin relaxation and glitches.

\section{STARQUAKES AND SPIN-DOWN OFFSETS}

The crust of a spinning neutron star is oblate with an equatorial bulge. The
moment of inertia of the bulge is
$I_{eb}
\sim I_{\rm crust} R^3
\Omega^2/(2 G M) \sim 4 \times 10^{-5} I_{\rm crust} \Omega_2^2 $ where $R\sim
10^6
$ cm is the stellar radius, $\Omega = 100 \Omega_2 $ rad s$^{-1}$ is the star's
spin frequency, and  $I_{\rm crust}$ is the  characteristic moment of inertia
of the crust. The crust contains about 1\% of the star's total moment of
inertia
$I_{\rm total}$.  As the star spins down, the equatorial circumference
shrinks and the polar radius grows. Because the crust is solid, strain
develops as the star's shape changes.  As sketched in  Figure 1, the
strain in the crust is relieved along starquake faults that form at an angle
to the star's equator \cite{lfe98, fle}.   Matter
slides along these faults to higher latitudes, and magnetic stresses favor
those faults that direct  matter along field lines  toward the magnetic
poles.

An important result of the earlier studies is that  starquakes shift the
stellar matter asymmetrically, creating excess  moment of inertia
$\delta I$   about an axis different from any of the pre-starquake principal
axes.  This distortion changes the 
orientation of the principal inertial axis by an angle
$\Delta \alpha \sim
\delta I/I_{eb} \sim 2.5 \times 10^4\Omega_2^{-2} \delta I/I_{\rm crust} =
2.5 \times 10^{-3}\Omega_2^{-2} \delta_{-7}$, where $\delta I/I_{\rm
crust}\equiv 10^{-7}
\delta_{-7}
$ \cite{lfe98}.  The {\it distortion parameter} $\delta_{-7}$
characterizes the
size of the starquake. 

 When the principal axis of inertia of the crust is not aligned with
the star's angular momentum, the star precesses and wobbles.
Eventually the star relaxes to a new equilibrium in which the axis and
angular momentum are again realigned, and the magnetic pole is shifted
by the angle $\Delta \alpha$ closer to the rotational equator.  In
some mechanisms for pulsar spin down, such as the magnetic dipole
braking model, this angular shift increases the torque on the star,
producing a long-lasting increase in the spin-down rate.  The
persistent spin-down offsets observed in the Crab pulsar can be
explained by this mechanism if $\Delta \alpha \sim 10^{-3}$
\cite{leb92le97} corresponding to $\delta_{-7} \simeq 1.6$ ( $\Omega_2
\simeq 2$ for the Crab pulsar).

\begin{figure}[ht]
\epsfig{file=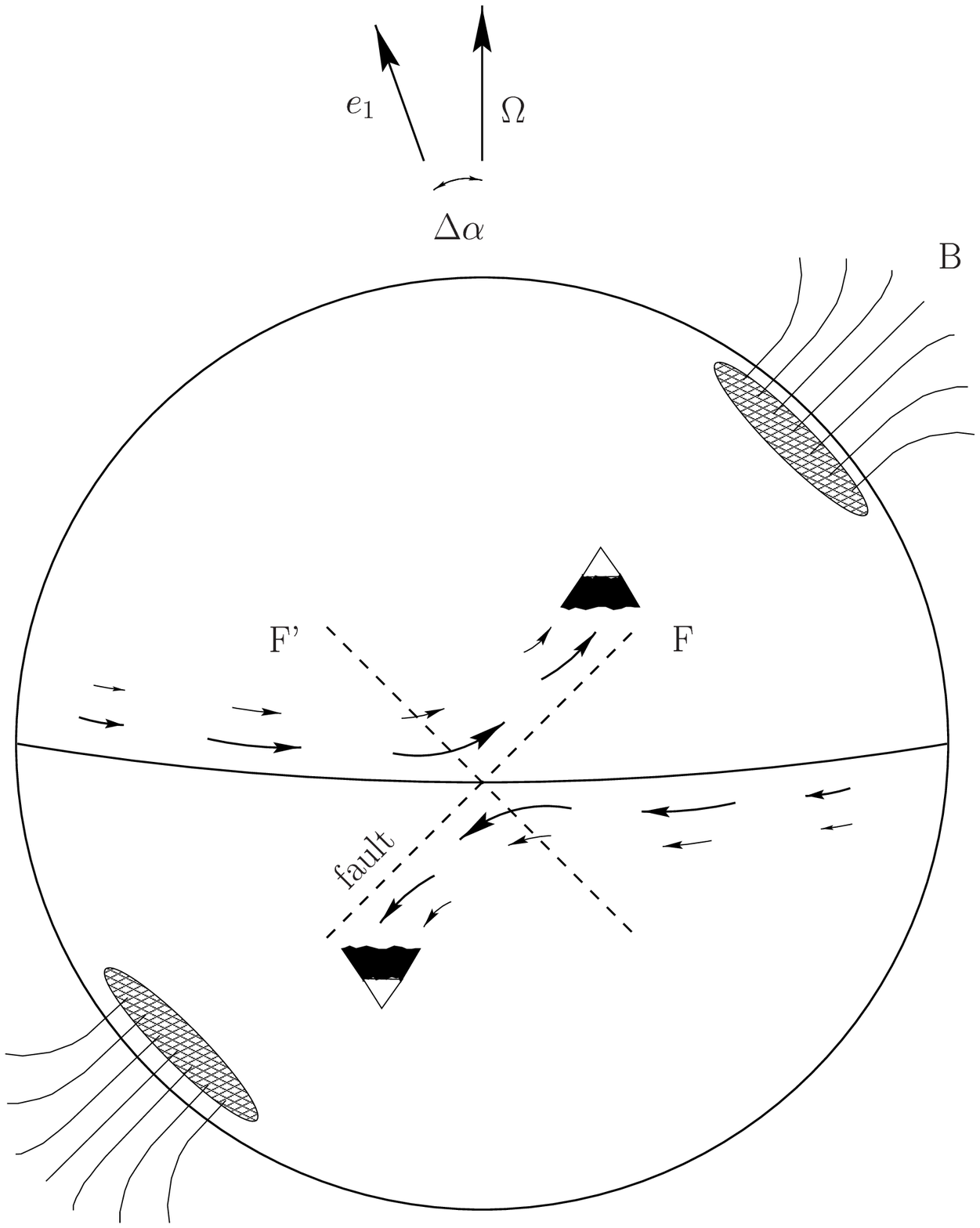,width=2.5in}
\vspace*{-2.5in}
\narrowcaption
{\protect\inx{{Starquakes} relieve the stress that builds as the a
neutron star's spin slows.   Matter can slide to higher latitudes along faults
$F$ or
$F^\prime$. Magnetic stresses favor faults such as $F$ that move matter toward
the magnetic poles.  The accumulated matter, shown as snow-capped peaks,
 shift the  principal axis of inertia by an angle $\Delta \alpha$ relative to
the star.} }
\end{figure}
\section{THE ORIGIN OF GLITCHES}

Starquake-induced asymmetry in the stellar crust excites precession.
If the neutron star crust behaved as an isolated rigid body, it would
precess or wobble at a frequency $\Omega_w \simeq (I_{eb}/I_{\rm
crust}) \Omega$ and the angle between the angular velocity and the
angular momentum would be $\sim (\Delta \alpha)^2 $.  The spin
behavior of a realistic neutron star is more complicated than this
for several reasons. First, the pinning of superfluid vortices in
the inner crust acts to stabilize the spin of the crust. Second, the
crust and core of the star are not strongly coupled on the precession
time scale. Third, the tilting of the crust accelerates vortex creep.

{\bf Pinned superfluid.} In the inner crust of the star the neutron
superfluid vortex lines may pin to the nuclei in the solid crust. The
rotation of the superfluid is determined by the location of the vortex
lines, and, as long as the vortex lines remained pinned, the
superfluid velocity field cannot change.  The gyroscopic action of
the pinned superfluid works with the equatorial bulge to further
stabilize the star \cite{s77swc}.  The moment of inertia $I_{\rm
pinned}$ of the pinned superfluid is comparable to that of the crust
and much larger than the moment of inertia of the equatorial bulge;
$I_{\rm pinned}\sim 0.01 I_{\rm total} \gg I_{eb}$. The pinned
superfluid decreases the equilibrium tilt of the star by  a factor
$I_{eb}/I_{\rm pinned}$ to an angle $\Delta \alpha_p \sim \delta
I/I_{\rm pinned} \sim 10^{-5} \delta_{-7} \ll \Delta \alpha$.  This is
the tilt angle of the star with completely pinned superfluid {\it
after the precession has damped} (the tilt immediately after the
starquake is $ \sim {[\Delta \alpha_p]}^2 \sim 10^{-10} \delta_{-7}^2
$).  The precession frequency is proportional to $I_{\rm pinned}$ and
inversely proportional to moment of inertia coupled to the crust. If
the spin of the core and the crust are tightly coupled, the star's
precession frequency is  $\Omega_w = (I_{\rm pinned}/I_{\rm total})
\Omega \simeq \Omega_2$ rad s$^{-1}$.

{\bf Coupling between the crust and
the core.} Changes in the crust's motion are communicated to the core by
MHD-like waves.  If the protons in the neutron star core form a type II
superconductor, as expected, the magnetic field is confined to thin  tubes of
flux $\pi \hbar c/e \sim 2
\times 10^{-7}$ G cm$^{-2}$, with characteristic dimensions  $\Lambda \sim 50 $
fm and  field strengths $B_\phi
\sim 10^{15}$ G. Signals travel from  the crust to the core on an
Alfv\`{e}n time
$t_{\rm couple} \sim (4 \pi \rho )^{1/2}  R/(B B_\phi )^{1/2} \sim 4
B_{12}^{-1/2}$ s, for an average field of $B = 10^{12} B_{12}$ G and a
density of
$\rho \sim 10^{15}$ g cm$^{-3}$ \cite{aeom98}. Because of the
imperfect coupling between the crust and the core, the precession is damped
on a
time scale \cite{bg}
\begin{equation}
t_{\rm damp} \simeq {\Omega\over\Omega_w } t_{\rm couple} \sim
{I_{\rm total}\over\ I_{\rm pinned}} t_{\rm couple} \sim 400\,
B_{12}^{-1/2} {\rm \
s.}\label{damp}
\end{equation}
The energy of the regular motion of
precession is converted into irregular fluid motions in the core.  The
irregular
motions of the superfluid neutrons and superconducting protons are then
converted into thermal energy by   processes such as electron scattering from
vortex  lines \cite{as}  or flux tube-vortex line interactions \cite{rzc}.
Since the core and
crust are not strongly coupled on the precession  time scale, the effective
moment of inertia of the precessing material is less, and the
precession frequency may
be higher than we have used. The damping time scale would be
correspondingly reduced.

After the precession has damped, the star's crust has tilted by
$\Delta \alpha_p$.  As the pinned vortex lines move to align with the
crust's rotation vector, the tilt angle grows until it equals $\Delta
\alpha. $ We now examine the processes by which the vortex lines move
and the effects of this motion on the star's spin behavior.

%\vskip2.0in
\begin{figure}[ht]
\vspace*{1.cm}
\hspace*{1.cm}
\epsfig{file=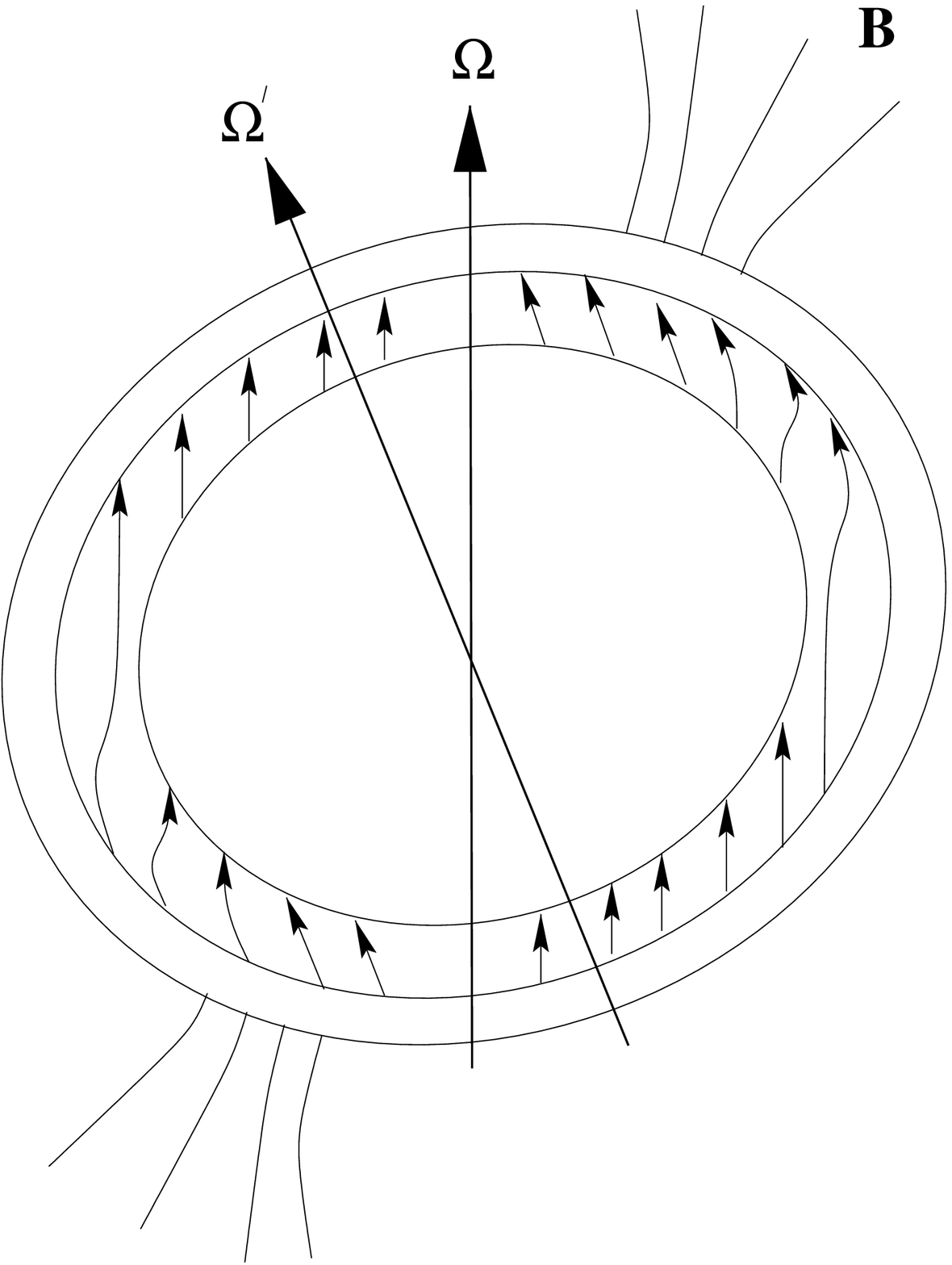,width=2.in}
\vspace*{-2.5in}
\narrowcaption{\protect\inx{{The shift} of the angular velocity from
$\Omega$ to $\Omega^\prime$ increases the velocity lag between the
superfluid and the solid crust in the upper right and lower left
sections of the star's inner crust in this sketch. In these sections
the vortex creep accelerates and the vortex lines migrate outward and
align with the new direction of the angular velocity.} }
\end{figure}

{\bf Motion of vortex lines.} In a steady state, the angular velocity of the
superfluid closely follows that of the solid crust.  As the superfluid rotation
slows, the vortex lines move radially  with a  steady velocity
$v_{ss} \simeq - R\dot
\Omega/  (2\Omega) = R/  (4 t_{\rm age}) $, where the spin-down age of the
pulsar is $t_{\rm age} \equiv \Omega /  (2 \dot \Omega)$ \cite{leb93}.
For the Crab
pulsar
$v_{ss}
\simeq 5
\times 10^{-6}$ cm s$^{-1}$ and for the Vela pulsar $v_{ss} \simeq 7 \times
10^{-7}$ cm s$^{-1}$.
If the vortex lines  move at $v_{ss}$ following a starquake, they would
align with the new direction of the crust's principal axis in a time $t_{\rm
align} \sim  \Delta \alpha R/v_{ss} \sim 10^{-2} \delta_{-7} \Omega_2^{-2}
t_{\rm
age}$, years for the Crab pulsar  and decades for
the Vela pulsar.
The vortex velocity can be greatly enhanced by the perturbations
produced by the tilt of the spin axis.\footnote{Heat produced during a
starquake can also increase the vortex velocity \cite{therm}.}
After the star's spin axis tilts   relative to the pre-starquake
orientation, some parts of the star are further from 
the new spin axis; this shift can be as large as
$\Delta \alpha_p R$. For these regions, the linear velocity from the  star's
spin increases by $\Delta \alpha_p R \Omega$. The superfluid velocity, on
the other
hand,  changes much less.  If the neutron vortex lines in the
star vortex were completely pinned, the superfluid velocity in the
star's frame would remain unchanged.  Even
though the superfluid in the core is not  pinned, the vortex lines cannot
move through the crust, so the crust superfluid velocity remains fixed.
The velocity lag
$v_\delta$ between the superfluid and the solid crust thus changes by as
much as
$
\Delta v_\delta \sim  \Delta \alpha_p R
\Omega \simeq  10^3 \delta_{-7}  {\rm \ cm\,s}^{-1}
$
 in parts of the star; see Figure 2. 
Velocity differences of this
magnitude may have dramatic effects on the vortex velocity, large enough to
produce the crustal spin-ups associated with glitches.

Pinned vortex lines in the inner crust can creep outward through
thermal activation of vortex segments from their pinning barrier;
a line segment unpins from one configuration and
migrates to new sites where it repins \cite{aaps, leb93}. The general
form of the creep velocity is $v_{\rm creep} \propto \exp[-A/kT]$,
where $A$ is the minimum activation energy for a segment of the vortex
line to unpin \cite{le91}. The activation energy depends sensitively
on the lag velocity $v_\delta$.  If the vortex lines are moving at
their steady state rate $v_{ss}$ and the lag velocity suddenly jumps
by $\Delta v_\delta$, the new creep rate is
\begin{equation}
v_{\rm creep} =v_{ss} \exp \left [ - {dA\over d v_\delta }{\Delta v_\delta
\over kT}\right ] .
\end{equation}
The activation energy  of a vortex line is
characterized by the maximum pinning force $F_{\rm pin}$ between a nucleus  and
a vortex line, the range of the pinning force $r_{\rm pin}$ and the distance
between pinning sites $\ell_{\rm pin}$. The maximum lag velocity between the
superfluid and the crust that the pinning forces can support is  $v_{\rm max}
=F_{\rm pin}/ \rho_s \kappa \ell_{\rm pin} $, where $\kappa = \pi
\hbar/(m_n)=2.0 \times 10^{-3} $
cm$^2$ s$^{-1}$ and $\rho_s$ is the mass density
of the neutron superfluid; $\rho_s \sim 10^{14}$ g
cm$^{-3}$  is the characteristic density of much of the inner crust
superfluid.  The activation energy required for a vortex line
to unpin depends  on the stiffness of  the vortex line; if the line is
flexible,
it can unpin from one nucleus at a time, whereas if it is stiff,
the line must unpin from many nuclei simultaneously \cite{le91}. The
parameter
$\tau\simeq
0.4 \rho_s \kappa^2 r_{\rm pin}/(F_{\rm pin} \ell_{\rm pin})$ characterizes
the
vortex line's stiffness \cite{le91}; for the conditions in much of the
crusts of
the Crab and Vela pulsars $\tau >1$ and vortex lines move by unpinning 
from many sites simultaneously. In this limit the appropriate expression for the
activation energy is
$
A \simeq 6.8 F_{\rm pin}r_{\rm pin}\tau^{1/2}(1-{v_\delta/ v_{\rm max}})
^{5/4},
$
and it's derivative is
\begin{equation}
 {dA\over d v_\delta }  \simeq - 4.8{ F_{\rm
pin}r_{\rm pin}\tau^{1/2}\over v_{\rm max}} \simeq - 3.0\left ( {
\rho_s^2 r_{\rm pin}^3 \kappa^3
\over v_{\rm max} } \right )^{1/2} .
\end{equation}
In obtaining this equation we set $ (1-{v_\delta/  v_{\rm
max}} )^{1/4} \sim 0.6$, which is a characteristic value for a variety of
stellar models \cite{leb93}.

We can use the observations of the Vela pulsar to estimate $v_{\rm
max}$.  This pulsar exhibited a string of 12 nearly evenly spaced
glitches separated with an average interval of $t_{\rm int} = 2.3$
years \cite{lel}.  The regularity of these glitches suggests that the
inner crust superfluid remains pinned between glitches until the
velocity lag between the solid crust and the superfluid approaches
$v_{\rm max}$. With this interpretation, a lower limit to the critical
lag velocity is $v_{\rm max}\gap \vert \dot \Omega_{\rm Vela} \vert
t_{\rm int} R = 7.0 \times 10^3 $ cm s$^{-1}$.  The value of $v_{\rm
max}$ exceeds this limit because the inner crust superfluid might
not relax to zero lag velocity at each glitch. For our estimates we
use $v_{\rm max}= 10^5 v_5 $ cm s$^{-1}$.  For $v_5 \simeq 1$ the
pinning force is $F_{\rm pin} \rho_s \kappa \ell_{\rm pin} v_{\rm max}
\simeq 13\, (\ell_{\rm pin}/100{\rm fm}) $ keV fm$^{-1}$ at a
superfluid density of $\rho_s = 10^{14}$ g cm$^{-3}$. This pinning
force is much smaller than that obtained by recent microscopic
calculations of vortex-nuclear interactions \cite{eb88pvb} assuming a
perfect crystal, but it is larger than the average pining force
estimated for an amorphous crust \cite{jones}.

Taking $\rho_s = 10^{14}$ g cm$^{-3}$ and interaction distance $r_{\rm pin} =
10$ fm \cite{nvde99}, the change in
the activation energy in parts of the crust after a starquake is
\begin{equation}
{dA\over d v_\delta }\Delta v_\delta  \sim - 0.053 v_5^{-1/2}\left( { \Delta
v_\delta
\over {\rm \ cm\,s}^{-1}}
\right){\rm
\ keV} \sim - 52 \, \delta_{-7} \Omega_{2} v_5^{-1/2}
{\rm \ keV} .
\end{equation}

The time scale for the post-starquake relaxation of the vortex lines is
$t_{\rm relax} \sim \Delta \alpha R/v_{\rm creep}$; for $kT$ in keV, this
gives
\begin{equation}
t_{\rm relax} \sim {\delta_{-7} t_{\rm age} \over 100 \Omega_2^2} \exp \left [
- {52 \, \delta_{-7} \Omega_2\over  v_5^{1/2} kT}\right ] .
\end{equation}
The vortex relaxation time is very sensitive to the magnitude of the
quake-induced shape change.  For example, taking $v_5=1$, we find that for the
Crab pulsar ($kT
\sim 20$  keV, $t_{\rm age} \sim 10^3 $ yr, $B_{12} \sim 4  $) the relaxation
time  is  less than the damping time if the distortion parameter is 
$\delta_{-7}
\gap 8.8$, but it is more than $1000 t_{\rm damp}$ if $\delta_{-7} \lap
4.3$.
The corresponding values for the distortion parameter for the Vela pulsar ($kT
\sim 8$  keV, $t_{\rm age} \sim 10^4 $ yr, $B_{12} \sim 8.8 $) are $\delta_{-7}
\gap 1.9$ and
$\delta_{-7} \lap 1.2$. For each pulsar, there is a critical value for
the distortion
parameter
$\delta_{-7}$. Starquakes that produce distortions  above this threshold
trigger rapid
vortex motions while  smaller events
generate only gradual changes.

{\bf Spin jumps.} In the large events, the rapid outward motion of the
vortex
lines produces a
 slowing of the superfluid and a corresponding spin up of the crust.
In the regions of the inner crust where $\Delta v_\delta$ is positive the
vortex
lines rapidly creep a distance $\Delta \alpha R \sim 2.5 \times 10^{-3}
\Omega_2^{-2}
\delta_{-7} R$. The superfluid in the affected regions of the inner crust
will now spin more slowly by  corresponding amount:
$\Delta \Omega_s/\Omega \sim 2.5 \times 10^{-3} \Omega_2^{-2}
\delta_{-7}$. The angular momentum lost by superfluid is imparted to the crust
and the core which is strongly coupled to it \cite{as}, giving  $I_{\rm
total} \Delta
\Omega_{\rm crust}= - I_{\rm unpinned} \Delta \Omega_s
$, where $I_{\rm unpinned}$ is the moment of inertia of the region of rapid
vortex creep.
 Since  only 1/2 of the crust has $\Delta
v_\delta> 0$, we have $I_{\rm unpinned}\lap 0.5I_{\rm crust}$ and
\begin{equation}
{\Delta \Omega_{\rm crust}\over \Omega}\sim - {I_{\rm crust}\over 2
I_{\rm total}} {\Delta
\Omega_s\over \Omega } \sim 1.3 \times 10^{-5}
{\delta_{-7}\over  \Omega_2^{2}}.
\end{equation}

The rapid creep of vortex lines following a large starquake could
explain the giant glitches with ${\Delta \Omega_{\rm crust}/ \Omega}
\sim 2 \times 10^{-6}$ observed in the Vela and other pulsars.

In the above estimate we assume that the lag between the
superfluid and the crust is large enough to  supply the needed angular
momentum to the crust. As the vortex lines creep through a distance $\sim
\Delta \alpha R$ the local superfluid slows by $\sim
\Delta \alpha R \Omega$.  If the pre-starquake lag velocity $v_\delta$
were less
than this value, the vortex creep would stop before the vortex lines move the
full distance
$\sim
\Delta \alpha R$.  A necessary condition for the above estimate to be
valid is that $v_\delta \gap \Delta \alpha \Omega R$.  Since $v_{\rm max}>
v_\delta$
we have $v_5\gap 2.5  \delta_{-7} \Omega_2^{-1}$.
The limiting factor in the size of a starquake-induced pulsar glitch may be
the pre-glitch lag velocity. The reason for the large difference between the
size of the Crab glitches and the giant glitches of the Vela pulsar may be that
in the Crab pulsar, with its higher internal temperature,  vortex creep
between glitches  limits the build up of a sufficiently large
$v_\delta$.

\section{SUMMARY}

Starquakes tilt the principal axis of inertia of a neutron star.  As the star
relaxes to its new equilibrium orientation, neutron superfluid vortex lines
migrate outward, spinning up the rest of the star.  The spin ups from large
quakes may explain the giant  glitches observed in isolated pulsars.  The
change in the direction of the magnetic axis may increase the spin-down torque,
as observed in the Crab and other pulsars.  Detailed calculations
of  the post-starquake relaxation for both large and  smaller events may yield
distinctive timing signatures to compare with observations.

%% optional
\begin{acknowledgments}

This work was performed under the auspices of the U. S. Department of Energy,
and was supported  by IGPP at LANL, NASA EPSCoR Grant  291748, NASA ATP Grants
NAG 53688 and NAG 52863, and by USDOE Grant DOE/DE-FG02-87ER-40317.

\end{acknowledgments}
%% appendix optional
%\appendix{This is the Appendix Title}
%This is an appendix with a title.

%\appendix{}
%This is an appendix without a title.

%
% Bibliography made with BibTeX:
%% apalike is preferred if you have used \kluwerbib, above.
%% Otherwise you may use any .bst style your editor approves.

\bibliographystyle{apalike}
%\chapbblname{<name of .bbl file>}
%\chapbibliography{<name of .bib file>}

%or

\end{document}